\documentclass[amsmath, reprint, floatfix]{revtex4-1}
\usepackage{txfonts}
\usepackage{xcolor}
\usepackage{microtype}
\usepackage{graphicx}

\begin{document}
\title{Infrared attosecond field transients and UV to IR few-femtosecond pulses generated by high-energy soliton self-compression}

\author{Christian Brahms}
\email[Corresponding author: ]{c.brahms@hw.ac.uk}
\author{Federico Belli}
\author{John C. Travers}
\affiliation{School of Engineering and Physical Sciences, Heriot-Watt University, Edinburgh, EH14 4AS, UK}

\begin{abstract}
Infrared femtosecond laser pulses are important tools both in strong-field physics, driving X-ray high-harmonic generation, and as the basis for widely tuneable, if inefficient, ultrafast sources in the visible and ultraviolet. Although anomalous material dispersion simplifies compression to few-cycle pulses, attosecond pulses in the infrared have remained out of reach. We demonstrate soliton self-compression of 1800~nm laser pulses in hollow capillary fibers to sub-cycle envelope duration (2 fs) with 27~GW peak power, corresponding to attosecond field transients. In the same system, we generate wavelength-tuneable few-femtosecond pulses from the ultraviolet (300~nm) to the infrared (740~nm) with energy up to 25~$\muup$J and efficiency up to 12\%, and experimentally characterize the generation dynamics in the time-frequency domain. A compact second stage generates multi-$\muup$J pulses from 210~nm to 700~nm using less than 200~$\muup$J of input energy. Our results significantly expand the toolkit available to ultrafast science.
\end{abstract}

\maketitle
\section{Introduction}
Ultrafast light sources in the infrared spectral region are important tools in optical science. Optical parametric amplifiers (OPA) in the infrared with subsequent frequency conversion stages can access the visible and ultraviolet spectral regions commonly used in ultrafast spectroscopy \cite{maiuri_ultrafast_2020}. These sources are widely tuneable, but suffer from low conversion efficiencies and cannot reach the few-femtosecond pulse durations required for experiments with cutting-edge time resolution. Few-cycle pulses in the spectral region around 1800~nm are particularly important in attosecond science as they enable the generation of soft X-ray attosecond pulses through high-harmonic generation \cite{cousin_high-flux_2014, johnson_high-flux_2018}. The most widely used method of compressing infrared pulses to few-cycle duration is nonlinear spectral broadening in gas-filled hollow capillary fibers (HCF) followed by phase compensation in bulk material \cite{schmidt_compression_2010}. However, the pulse duration achievable with this method is limited by uncompensated higher-order dispersion \cite{austin_spatio-temporal_2016,schmidt_cep_2011}. Much like in HCF-based pulse compression at other wavelengths, the dispersion of the waveguide itself is usually ignored.

Optical soliton dynamics, based on the interplay between nonlinearity and dispersion during propagation of an ultrafast laser pulse, can be used for both pulse compression \cite{mollenauer_experimental_1980} and frequency conversion through resonant dispersive wave (RDW) emission \cite{wai_nonlinear_1986, im_high-power_2010}. This is particularly powerful in gas-filled hollow-core waveguides due to their large guidance bandwidth, excellent power handling as well as pressure-tuneable nonlinearity and dispersion. For low pulse energies (few $\muup$J), anti-resonant photonic crystal fibers (AR-PCF) offer low-loss guidance with small core diameters (typically around $30$~$\muup$m), and soliton effects have been demonstrated at a variety of driving wavelengths \cite{cassataro_generation_2017,balciunas_strong-field_2015,kottig_generation_2017,joly_bright_2011,hosseini_uv_2018}. Recently, we have shown that soliton dynamics can be observed in HCF at much higher energies by employing pre-compressed driving pulses \cite{travers_high-energy_2019,brahms_high-energy_2019}. In numerical studies, moderate self-compression at 1800~nm \cite{zhao_self-compression_2017} as well as sub-cycle pulse generation in the mid-infrared \cite{voronin_subcycle_2014} have previously been predicted, however no experiments have been performed to date.

Here, we demonstrate soliton self-compression of 30~fs pulses at 1800~nm to 2~fs, corresponding to 840~attosecond field transients, in a gas-filled hollow capillary fiber. Due to the stronger HCF dispersion at longer wavelengths, this is achieved in a single stage without pre-compression. The formation of attosecond pulses at significantly lower photon energy than demonstrated previously dramatically expands the toolset available for studies in strong-field physics and beyond. Through RDW emission, this source is also capable of single-stage tuneable frequency conversion from 1800~nm to highly energetic few-femtosecond pulses from the ultraviolet (UV) to the near infrared, continuously covering the entirety of the visible spectral range. Using the self-compressed pulses to drive a second HCF stage, we demonstrate more widely tuneable RDW emission in a much more compact setup. By also reducing the energy requirements, this makes our frequency conversion scheme widely accessible and allows it to be integrated into multi-color time-resolved spectroscopy experiments. Finally, we directly study the dynamics underlying the self-compression and RDW emission without the aid of numerical simulations by employing an ultrabroadband pulse characterization technique.

\section{Results}
\begin{figure}
    \centering
    \includegraphics[width=3.39in]{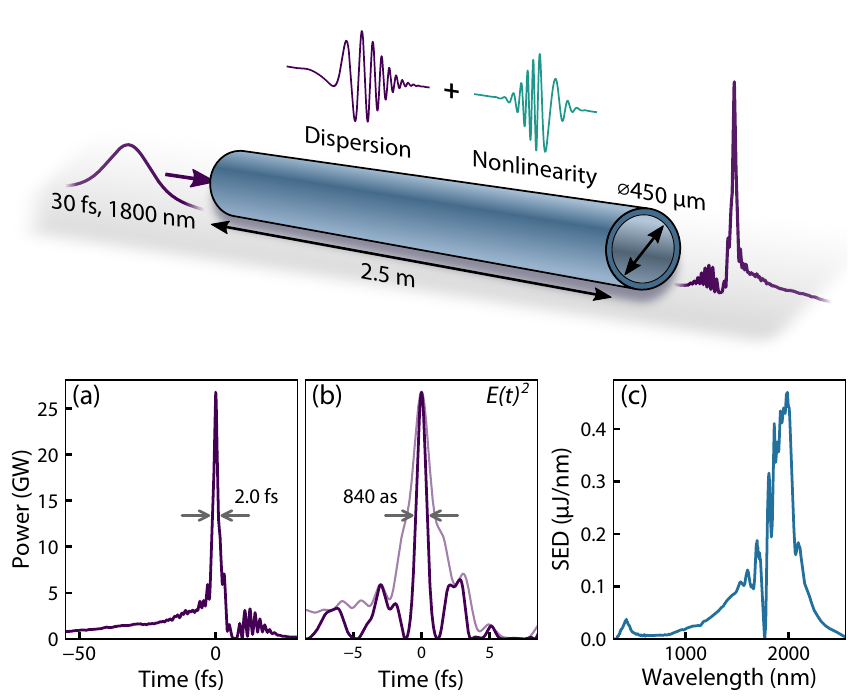}
    \caption{{Self-compression to infrared attosecond pulses.} Laser pulses at 1800~nm central wavelength with a duration of 30~fs FWHM are coupled into a 2.5~m long gas-filled HCF with an inner diameter of 450~$\muup$m. In the HCF, the combination of anomalous dispersion arising from the waveguide and positive nonlinearity provided by the gas leads to self-compression of the pulses. With an argon pressure of 470~mbar and an initial pulse energy of 289~$\muup$J, the pulse at the HCF exit has an envelope duration of 2~fs FWHM (a), which corresponds to an 840~as field transient (b). The spectrum of this pulse extends from the ultraviolet (400~nm) to the mid-infrared (2550~nm) (c).}
    \label{fig:setup}
\end{figure}

\begin{figure*}
    \centering
    \includegraphics[width=7.05in]{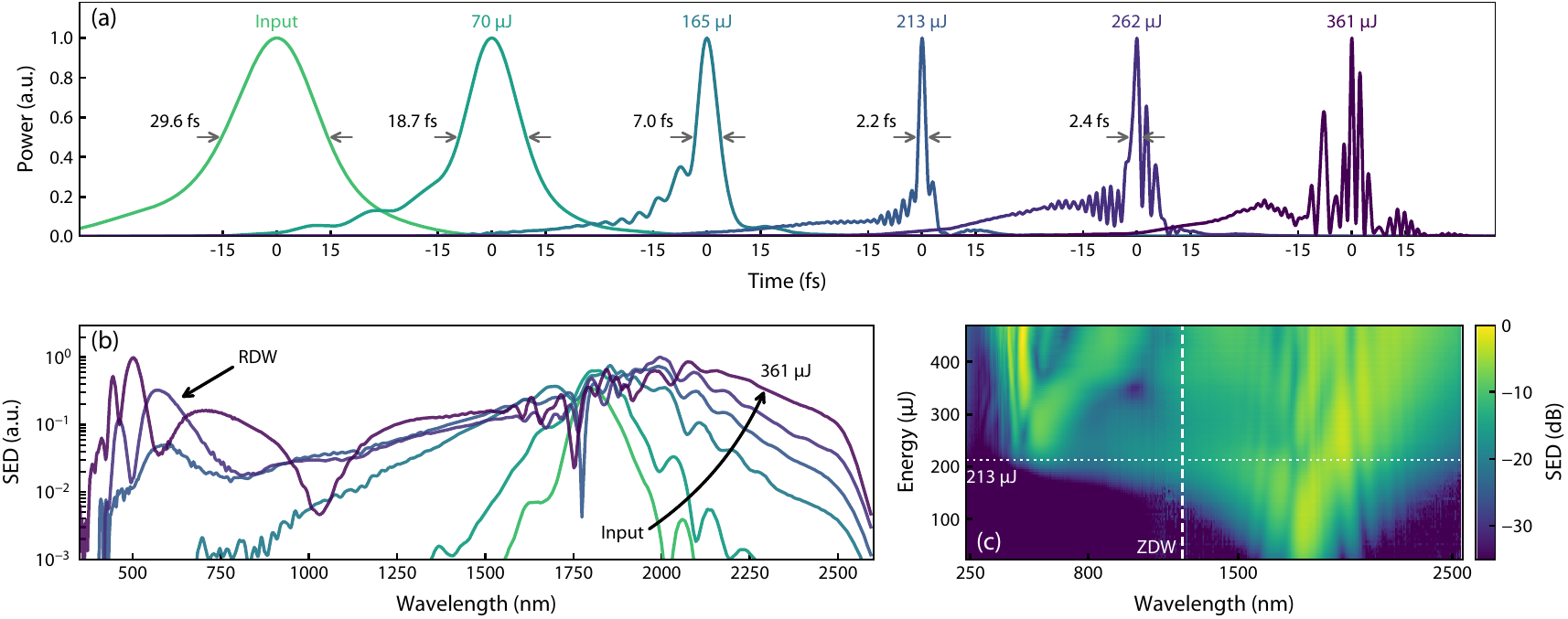}
    \caption{{Measured dynamics of the self-compression process.} (a) Reconstructed pulse profiles at the exit of the 450~$\muup$m HCF when filled with 830~mbar of argon as measured using TDP. The first line shows the input pulse (measured with the HCF evacuated) and each subsequent line the pulse profile when the HCF is pumped with the energy indicated. Each pulse profile is normalized to its peak. The labels give the FWHM duration of the pulses. (b) Power spectrum of the pulses presented in (a). (c) Output power spectrum as a function of input energy for the same conditions on a logarithmic color scale. The horizontal dotted line marks the point of maximal self-compression, corresponding to the 2.2~fs pulse shown in (a). The vertical dashed line marks the zero-dispersion wavelength.}
    \label{fig:compression}
\end{figure*}

\subsection{Sub-cycle infrared self-compression}
Figure~\ref{fig:setup} summarizes the principle of the experiment. To obtain soliton self-compression, we couple pulses with a duration of 30~fs (all pulse durations will be given as the full width at half maximum (FWHM) of the power) and a central wavelength of 1800~nm into a 2.5~m long, 450~$\muup$m diameter HCF filled with argon. The pulses are characterized spectrally by a calibrated spectrometer and temporally using ultrabroadband time-domain ptychography (TDP) \cite{witting_time-domain_2016}. From these measurements, the pulse exiting the HCF is reconstructed by numerical back-propagation (for more details see section \ref{sec:methods}).

Soliton dynamics are based on the simultaneous action of a positive nonlinearity (in this case, the Kerr effect in the filling gas) and negative second-order dispersion (provided by the HCF for sufficiently low filling gas pressure). The balance between these two influences determines the evolution of the laser pulse. In our case, the nonlinearity initially dominates, leading to spectral broadening, which then increases the importance of dispersion. The positive frequency chirp induced by the Kerr effect is thus partially compensated and the pulse self-compresses, in turn enhancing the nonlinearity. This feedback process is arrested only when the pulse spectrum has expanded so dramatically that dispersion dominates instead.

Figure~\ref{fig:setup}(a) shows the measured profile of a 289~$\muup$J, 30~fs pulse after undergoing self-compression in the HCF when filled with 470~mbar of argon (these parameters correspond to a soliton order of 4.5 and a zero-dispersion wavelength (ZDW) of 1066~nm \cite{travers_high-energy_2019}). The pulse has self-compressed to 2~fs, one third of one cycle of the driving field (6~fs at 1800~nm). A pedestal at lower power remains and precedes the pulse. The self-compressed pulse contains 211~$\muup$J of energy and its peak power is 27~GW. As shown in Fig.~\ref{fig:setup}(b), the electric field transient created here is significantly shorter than the envelope, with a FWHM duration of the square of the field of only 840~as. Such a short duration requires a higher frequency than that of the driving field---the central wavelength of the peak itself (calculated as the first moment of the spectrum after applying a window function around the peak) is shifted to 1490~nm [see Fig.~\ref{fig:setup}(c)]. The measured and retrieved TDP traces are shown in Supplementary Fig.~S1.

In Fig.~\ref{fig:compression}, we investigate the self-compression dynamics in more detail. Figure~\ref{fig:compression}(a) shows the measured temporal evolution of the pulse exiting the HCF as the input energy is increased for a higher argon pressure of 830~mbar. From the initial duration of around 30~fs, the pulse progressively compresses towards the sub-cycle regime. At low input energy (70~$\muup$J), the pulse has compressed to 18.7~fs by the end of the HCF. Since the nonlinearity dominates during the initial propagation, this pulse is positively chirped with a Fourier-transform limit (FTL) of 16.3~fs. Increasing the input energy to 165~$\muup$J causes further compression to 7~fs with an FTL of 6.1~fs.

The pulse reaches its minimum duration of 2.2~fs at an input energy of 213~$\muup$J. The FTL at this point is 2~fs, but the central peak is virtually free of chirp. At even higher energy, the point of self-compression is significantly before the end of the HCF. This enables efficient RDW emission, in this case at around 500~nm, as shown in Fig.~\ref{fig:compression}(b) and (c). The complex structure in the pulse profile at 262 and 361~$\muup$J is caused by the interference between the various spectral components, which rapidly dephase after the self-compression point because of higher-order dispersion. The measured and retrieved TDP traces are shown in Supplementary Figures S2 and S3.

One notable feature of the self-compression to sub-cycle pulse duration is the extreme self-steepening of the pulse. In the time domain, self-steepening delays the peak of the pulse relative to the weaker pedestal, so that the peak appears at the trailing edge of the pulse. This is also visible in the strong asymmetry of the output power spectrum, which enhances the efficiency of RDW emission \cite{joly_bright_2011} and shifts the central wavelength (the first moment of the whole spectrum) of the self-compressed pulse to around 1640~nm. Note that since each pulse in Fig.~\ref{fig:compression}(a) is normalized to its peak, the pedestal appears more intense after the self-compression point at 213~$\muup$J.

Figure~\ref{fig:compression}(c) shows the spectrum of the pulse as a function of driving pulse energy. The spectrum shows the dramatic asymmetry in the spectral expansion accompanying self-compression. At its widest, the supercontinuum extends from 370~nm to beyond 2600~nm at the $-30$~dB level. The resonant dispersive wave at around 500~nm first appears at the point of maximum self-compression (213~$\muup$J, indicated by the white dotted line) and becomes more intense as the energy is increased. The white dashed line shows the zero-dispersion wavelength, which at 830~mbar pressure is 1240~nm.

\subsection{UV to infrared dispersive wave emission}
Resonant dispersive waves are emitted as a consequence of higher-order dispersion acting on the extremely broadband spectrum of the self-compressed soliton \cite{wai_nonlinear_1986, akhmediev_cherenkov_1995, karpman_radiation_1993, erkintalo_cascaded_2012}. Phase-matching between the soliton and a spectral band in the normal-dispersion region enables coherent build-up of energy far away from the frequency of the initial driving pulse. The phase-matching region can be broad, and the length over which the RDW interacts strongly with the soliton is limited by pulse breakup of the latter. As a consequence, the pulses generated in this manner can be very short. Driving at 800~nm, RDWs have previously been tuned from the vacuum ultraviolet (110~nm) \cite{ermolov_supercontinuum_2015, travers_high-energy_2019} to the visible (550~nm) \cite{mak_tunable_2013}. By using a driving pulse at 1800~nm, we expand the tuneability of the RDW to longer wavelengths, so that the range extends from the ultraviolet to the near infrared.

\begin{figure*}
    \centering
    \includegraphics[width=7.05in]{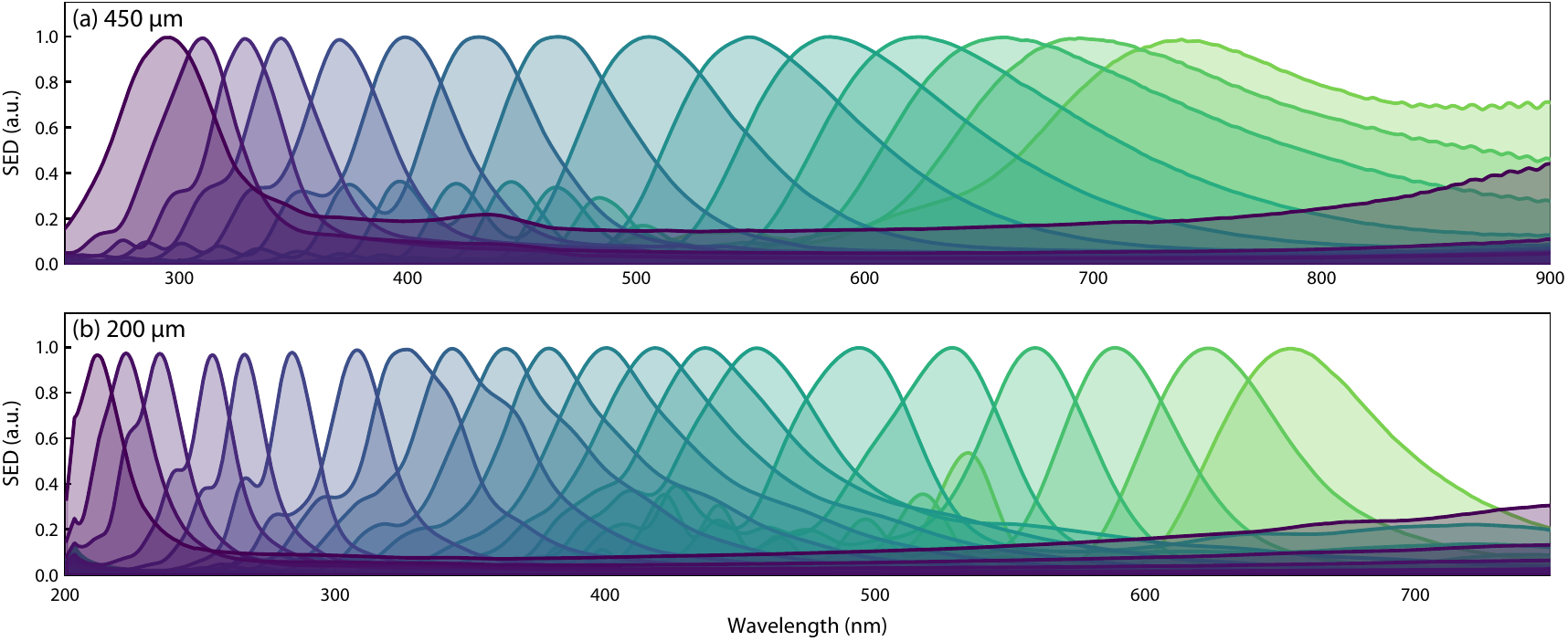}
    \caption{{Generation of tuneable resonant dispersive waves from the ultraviolet to near infrared.} (a) RDW spectra generated in the first HCF with 450~$\muup$m core diameter. Each spectrum is normalized to the peak in the wavelength region shown. The argon pressure required for the generation ranges from 255~mbar (shortest wavelength) to 1330~mbar (longest wavelength).
    (b) RDW spectra generated in the second HCF (200~$\muup$m core diameter) when driven with 16~fs pulses generated in the first HCF. The pressure varies from 470~mbar (shortest wavelength) to 7000~mbar (longest wavelength).}
    \label{fig:pres_scan}
\end{figure*}

\begin{figure}[b]
    \centering
    \includegraphics[width=3.39in]{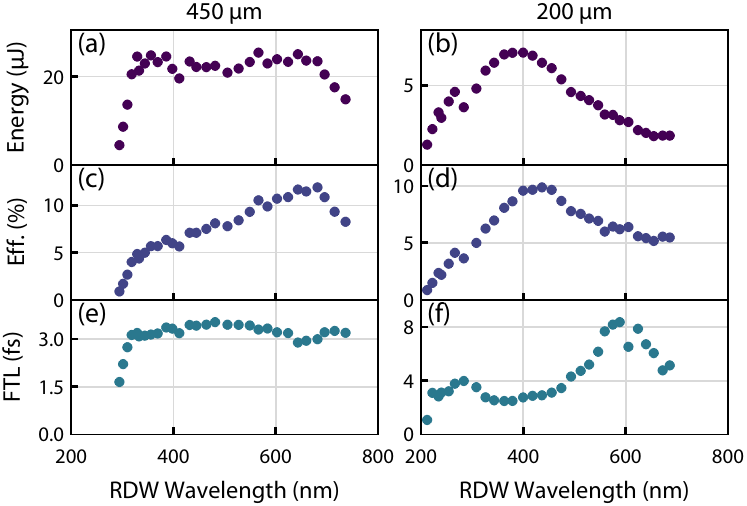}
    \caption{{Performance of the tuneable ultraviolet to infrared frequency conversion.} (a, b) Energy, (c, d) conversion efficiency, and (e, f) FTL duration of the RDWs shown in Fig.~\ref{fig:pres_scan}.}
    \label{fig:performance}
\end{figure}

Figure~\ref{fig:pres_scan}(a) shows RDW spectra obtained by changing the argon pressure in the 450~$\muup$m HCF from 255~mbar to 1330~mbar and tuning the input energy such that RDW emission occurs close to the end of the HCF. The peak wavelength of the RDW changes from 300~nm to 740~nm. The energy converted to the RDW is shown in Fig.~\ref{fig:performance}(a). It is calculated by applying a window function around the RDW peak and integrating the resulting spectrum. At the shortest wavelengths (lowest pressures), the energy increases sharply as the pressure is increased, since the generation is limited by the available driving pulse energy. The converted energy remains around 25~$\muup$J from a wavelength of 330~nm until around 700~nm before decreasing. The conversion efficiency, estimated from the input energy, coupling efficiency and RDW energy, is shown in Fig.~\ref{fig:performance}(c). It is around 1\% for the shortest wavelengths and increases steadily to 12\% in the near infrared.

Figure~\ref{fig:performance}(e) shows the transform-limited duration of the dispersive wave. It is extracted by Fourier transforming the filtered RDW spectrum and calculating the FWHM duration of the resulting pulse. The frequency bandwidth of the RDW is nearly constant over most of the tuning range, with an FTL duration of around 3~fs, except at the shortest wavelengths, where the RDW generation is not fully saturated. The shape of the RDW spectrum at 740~nm clearly shows the limits of the tuneability. At 1330~mbar pressure, the dispersion at the driving pulse wavelength is relatively weak (the ZDW is 1394~nm) and the dispersion landscape in the near infrared very flat, so that the RDW blends into a continuum which spans 600~nm to 2500~nm instead of emerging as an isolated spectral peak.

With shorter driving pulses, soliton self-compression occurs over shorter distances, allowing the use of smaller core sizes and lower overall energy as well as making the system more compact \cite{brahms_high-energy_2019}. We apply this principle to demonstrate that tuneable RDW emission from the ultraviolet to near infrared can also be obtained with driving pulse energies of less than 200~$\muup$J in a compact system.

Spectra obtained in a second HCF with a length of 38~cm and a core diameter of 200~$\muup$m are shown in Fig.~\ref{fig:pres_scan}(b). The driving pulses for this system were pre-compressed to 16~fs in the first HCF. As with the larger core size, the RDW can be tuned throughout the visible spectral range and to the near infrared (700~nm). The smaller core size increases the achievable intensity, which extends the tuning range to the deep UV (210~nm). With the RDW at its shortest wavelength, the supercontinuum generated here covers the entirety of our detection window, 200 to 2600~nm, at the $-30$~dB level. The converted energy is shown in Fig.~\ref{fig:performance}(b). Starting from the shortest wavelengths, it rises slowly from 1.5 to 7.5~$\muup$J up to 400~nm wavelength, after which it begins to drop.

We estimate the conversion efficiency, shown in Fig.~\ref{fig:performance}(d), as 1\% for the RDW at 210~nm, rising to 10\% at 450~nm and then dropping again to 5.5\% at 700~nm. Note that these efficiencies once again exclude the coupling loss of approximately 15\%---the total input energy to the second HCF (before coupling losses) ranges from 41~$\muup$J for infrared RDW emission to 185~$\muup$J in the ultraviolet. We observe a significant drop in the total transmission of the HCF system at the shortest RDW wavelengths (from 44\% at high pressure to 37\%). This suggests that the drop in RDW energy at short wavelengths is due to ionization losses resulting from the higher intensity. These could be largely avoided by switching to a gas with a higher ionization potential, such as neon or helium. At the longer wavelengths, the decrease in energy is more likely due to the bandwidth of the initial pulse. The spectrum of the self-compressed driving pulse already extends to around 1300~nm, which is close to or in the normal dispersion region (the ZDW at 7000~mbar is 1410~nm). As a consequence, the driving pulse energy cannot be increased very far before the spectrum forms a  continuum without a distinct RDW peak, which in turn reduces the measured RDW energy.

\subsection{Time-frequency measurements of soliton dynamics}
\begin{figure*}
    \centering
    \includegraphics[width=7.05in]{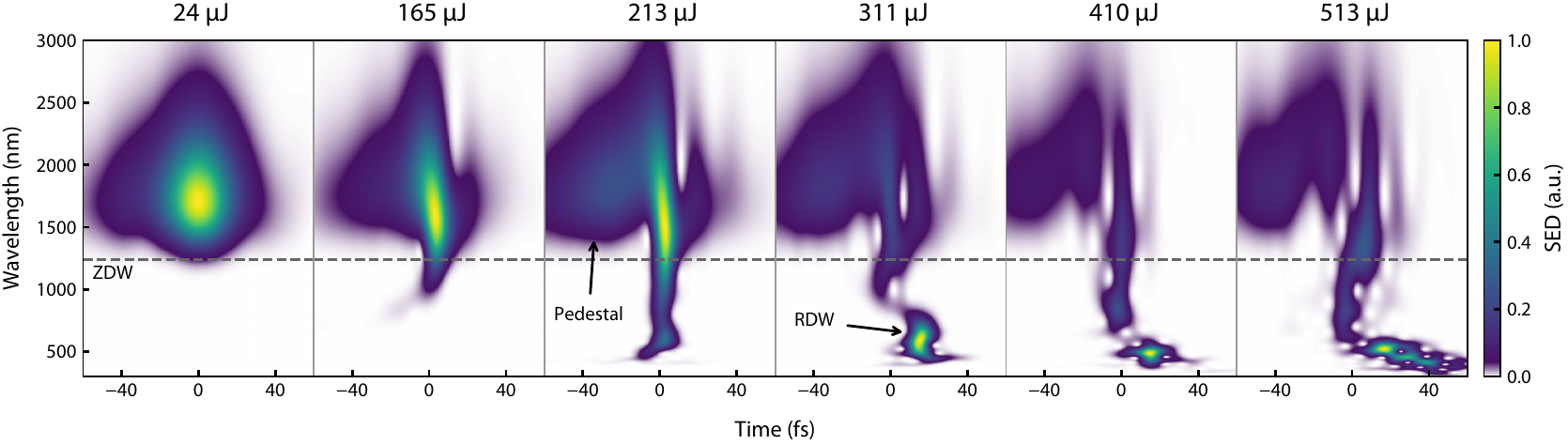}
    \caption{{Time-frequency measurements of self-compression and RDW emission.} Each column presents the spectrogram obtained of the pulse exiting the HCF, reconstructed from the TDP measurements through numerical back-propagation. The driving pulse energy is indicated above each spectrogram. The argon pressure was 830~mbar. The gate used was Gaussian with a FWHM of 10~fs. The color scale for each plot is normalized to the peak.}
    \label{fig:spectrograms}
\end{figure*}
Our ultrabroadband pulse characterization allows us to directly measure the dynamics of RDW emission in the time-frequency domain. Figure~\ref{fig:spectrograms} shows the spectrogram of the field exiting the 450~$\muup$m HCF when filled with 830~mbar of argon (the same conditions as shown in Fig.~\ref{fig:compression}) for different input pulse energies. This provides an approximate picture of the pulse evolution along the HCF, because the point of self-compression moves along the waveguide as the energy is changed.

The first three traces in Fig.~\ref{fig:spectrograms} show the self-compression from 30~fs to 2.2~fs. At 213~$\muup$J, the main pulse is fully compressed and free of chirp, as evidenced by the absence of tilt or curvature in the spectrogram. The RDW first appears at the self-compression point, since this is where the pulse spectrum first extends to the phase-matched wavelength. Energy continues to be converted as the main pulse breaks up, and at 311~$\muup$J the RDW is at its most broadband and shortest duration (3.7~fs). At 410~$\muup$J, the RDW has moved away from the infrared pulse due to the difference in group velocity, and conversion has ceased. The bandwidth at this point is reduced---despite being mostly free of chirp, the RDW is now 4.2~fs in duration. Subsequently, it evolves largely without nonlinear interaction, stretching in time due to group-velocity dispersion. For experiments requiring the RDW to be compressed, a pressure gradient can be used to deliver the pulse straight to vacuum. Measurements using AR-PCF have shown that the RDW can be preserved as a near transform-limited pulse using this approach \cite{brahms_direct_2019}.

As the energy is increased from 213~$\muup$J to 410~$\muup$J, the central wavelength of the RDW changes from 600~nm to 490~nm. This effect is due to the nonlinear contribution to the RDW phase-matching \cite{joly_bright_2011}, which becomes more important for larger driving pulse energy. The frequency shift is one key aspect in which varying the energy is not directly analogous to following the propagation dynamics along the waveguide---for a constant gas pressure and energy, the RDW does not shift significantly during propagation after it is generated.

The spectrograms also reveal the origin of the pedestal preceding the self-compressed pulses shown in Figs.~\ref{fig:setup} and \ref{fig:compression}. It is formed because the peak of the initial pulse experiences the strongest self-compression, while the tails remain relatively unperturbed---this is a well-known effect in soliton self-compression \cite{chen_nonlinear_2002}. The intensity dependence of the group velocity (i.e. self-steepening) causes the peak to fall behind, resulting in the pedestal on the leading edge. Due to its low intensity, it does not interact nonlinearly by itself. Instead, it accumulates anomalous group-delay dispersion, visible in the tilt in the spectrogram.

\section{Discussion}
Driven by applications, the generation of ever shorter laser pulses has been a major aim in ultrafast optics for several decades. Optical attosecond pulses, first generated using a light-field synthesizer \cite{hassan_optical_2016} and later by soliton self-compression in HCF \cite{travers_high-energy_2019}, represent the limiting case of this evolution in the visible spectral region. With an envelope duration below the cycle time of the field, the duration of the electric field transient itself becomes the relevant quantity. The self-compressed pulses we have demonstrated here extend this concept to the infrared, where such extremely short pulses have not previously existed. Low-energy infrared sub-cycle transients have been created by the interference of a few-cycle pulse with its second harmonic \cite{hammond_producing_2017}, but neither further compression to attosecond duration or energy scaling have been demonstrated with this technique. The new capability of generating infrared attosecond pulses we demonstrate here holds great promise, for instance in the study of solid-state strong-field physics \cite{ghimire_high-harmonic_2019}. The compression relies on neither special optics, such as dispersive mirrors, nor material properties, such as anomalous bulk dispersion. Therefore, changing the central wavelength of the driving pulse will allow for fine-tuning of the width of the transient.

Being based on back-propagation, the measurement is sensitive to the experimental error on the propagation distances. The overall effect of this error is less pronounced in the infrared as compared to the visible, because most materials are less dispersive in this spectral region. The envelope duration of the self-compressed pulse shown in Fig.\ref{fig:setup} is affected significantly and increases from 2~fs to 2.9~fs within the estimated uncertainty. However, the width of the field transient increases by less than 20\%, and an error in the back-propagation merely reduces the contrast between the main peak and its most intense satellite from a factor of 4 to a factor of 2.5 (see Supplementary Fig.~S4).

For HCF-based soliton self-compression at 800~nm, pre-compression of the driving pulse is useful in keeping the required HCF length manageable \cite{travers_high-energy_2019}. It is not necessary in the present experiment---significantly simplifying the setup---due to the wavelength dependence of the waveguide dispersion. The length scale of soliton self-compression is largely determined by the pulse duration and the dispersion at the driving wavelength, with shorter pulses and larger (more anomalous) dispersion leading to more rapid self-compression \cite{travers_high-energy_2019,brahms_high-energy_2019}. While the HCF has to be larger for longer driving wavelengths to compensate for increased loss (the propagation loss scales with the wavelength $\lambda$ and the core radius $a$ as $\lambda^2/a^3$), nevertheless the anomalous dispersion contribution of the waveguide is stronger (it scales as $\lambda^3/a^2$). The resulting reduction in required length enables soliton self-compression in a single HCF stage.

Another effect of the increased driving wavelength is the greatly extended tuneability of the dispersive wave to the infrared. A source that covers the entirety of the spectral range from ultraviolet to near infrared with a single frequency conversion stage is an extremely powerful tool for spectroscopy. Two additional features make this source particularly useful. Firstly, since HCF is free of the guidance resonances intrinsic to AR-PCF \cite{tani_effect_2018}, this range is covered entirely without gaps. Secondly, the conversion efficiency is high despite the very large frequency shift. Even at the shortest wavelength generated here---which corresponds to the 9$^\mathrm{th}$ harmonic of the driving field---around 1\% of the input energy is converted. Preliminary investigations show that by fully optimising the pre-compressed driving pulse and switching to a different gas with higher ionization potential, RDW emission in the vacuum ultraviolet, well below 200~nm, should also be possible in this system.

Recent progress in high-harmonic generation (HHG) using infrared drivers has enabled the first time-resolved X-ray absorption spectroscopy experiments using table-top sources \cite{pertot_time-resolved_2017} as well as the generation of attosecond pulses in the X-ray region \cite{cousin_high-flux_2014, johnson_high-flux_2018, cousin_attosecond_2017}. Tuneable few-femtosecond pump pulses derived from the same infrared pulse that drives HHG would enable pump-probe spectroscopy in a variety of systems while keeping the timing jitter between pump and probe to a minimum. However, the extremely low conversion efficiency in HHG limits the pulse energy available to drive any other frequency conversion. As we have demonstrated, widely tuneable RDWs can be generated in a compact setup driven by only a fraction of the typical output energy of infrared pulse compressors \cite{johnson_high-flux_2018, cousin_attosecond_2017, austin_spatio-temporal_2016}. The reduced energy requirements, continuous tuneability and short pulse duration make RDW emission driven by few-cycle infrared pulses an ideal source for widely applicable X-ray transient absorption experiments. This approach also makes UV to near-infrared RDW emission sources widely accessible even with lower-energy laser sources.

As with soliton effects in gas-filled waveguides in general, the dynamics underlying the source demonstrated here can be up-scaled in energy by increasing the HCF core size and length and decreasing the gas pressure \cite{travers_high-energy_2019,heyl_scale-invariant_2016}. The system used here is designed to demonstrate a wide range of tuneability as well as the complete dynamics of self-compression and RDW emission, so it is not optimized for maximum input energy (the total energy throughput of the optical components before the first HCF is around 55\%). Using the whole of the available energy at 1800~nm without attenuation, the correct choice of the HCF length and gas pressure would enable the generation of much more energetic sub-cycle pulses and RDWs at the cost of reduced RDW wavelength tuneability.

Finally, we note that the generation of single-cycle (4.5~fs) infrared pulses on target (i.e. without back-propagation) has previously been achieved using soliton dynamics in AR-PCF, albeit at more than one order of magnitude lower energy \cite{balciunas_strong-field_2015}. In our experiments, the achievable pulse duration at the interaction point of our characterization setup is limited to around 7~fs by the air path required. Minimising this---most effectively by delivering the pulses directly to vacuum---would allow for significantly shorter pulses to be obtained. In combination with the energy optimization discussed above, this provides a route towards millijoule-scale sub-cycle infrared pulses---an ideal driver for X-ray HHG.

In summary, we have demonstrated soliton self-compression of infrared laser pulses to sub-cycle duration at high pulse energy, as well as highly energetic, widely tuneable dispersive wave emission in simple hollow capillary fibers. We obtained up to 27~GW of peak power in a sub-cycle (2~fs) pulse at 1490~nm, extending the concept of optical attosecond pulses to longer wavelengths and providing a unique tool for ultrafast science. In the same, single-stage system, RDW emission enables the generation of ultrashort pulses which can be tuned from the ultraviolet, across the entire visible and to the infrared spectral range, carrying up to tens of $\muup$J of energy. A second HCF stage, pumped by the self-compressed pulses from the first, allows generation at even shorter wavelengths in a more compact system and with less input energy. Both systems are capable of generating supercontinua spanning over three octaves from the ultraviolet through most of the near infrared. With extremely broadband pulse characterization, we have further been able to experimentally investigate the self-compression and RDW emission dynamics in a hollow-core waveguide in detail. With the emergence of new ultrafast pump sources in the infrared, such as optical parametric chirped-pulse amplification systems and thulium-doped fiber lasers, we expect that our work will form the foundation for a new generation of ultrafast pulse compression and frequency-conversion technology.

\section{Methods}
\label{sec:methods}
\subsection*{Experimental apparatus}
A titanium-doped sapphire laser amplifier (Coherent Legend Elite Duo USX) delivers pulses of 25~fs duration and 8~mJ energy at 800~nm wavelength and a repetition rate of 1~kHz, which are then converted to 30~fs pulses at 1800~nm wavelength with 1.2~mJ energy in a white-light seeded OPA (Light Conversion HE-TOPAS Prime). To compensate for the anomalous group-velocity dispersion of other optical components, the pulses pass through a 5~mm thick piece of potassium bromide (KBr) followed by silica wedges for dispersion fine-tuning. An achromatic half-wave plate and Brewster-angle silicon plate form a variable attenuator. The pulses are then coupled into a 2.5~m long HCF with 450~$\muup$m core diameter. The HCF is stretched in order to eliminate bend loss \cite{nagy_flexible_2008} and mounted in a gas cell which is filled with argon.

Self-compressed driving pulses for the compact RDW emission system are generated by reducing the pressure in the first HCF to 180~mbar, delivering 16~fs pulses at the input to the second stage. These pulses then pass through another attenuator and are re-focused into a second HCF with 200~$\muup$m core diameter and 38~cm length, which is filled with argon at pressures between 470 and 7000~mbar.

A detailed sketch of the experimental apparatus is shown in Supplementary Fig.~S5.

\subsection*{Calibrated spectral characterization}
For spectral characterization, we use a combination of an integrating sphere and two fiber-coupled spectrometers, one for the range 200~nm to 1100~nm (Avantes ULS2048XL) and one for 1000~nm to 2600~nm (Avantes NIR256-2.5). This system is calibrated as a whole for absolute spectral response, so that spectral energy density and pulse energy can be extracted directly from the spectra. We additionally cross-calibrate by comparing the total energy in the spectrum to that measured by a calibrated thermal power meter. The use of an integrating sphere to collect the whole output beam is critical, since the output spectrum can extend over more than 3 octaves, corresponding to a change in beam area of nearly a factor of 100 between the edges of the spectrum.

\subsection*{Pulse characterization}
We characterize the self-compressed pulses using sum-frequency generation time-domain ptychography (TDP) \cite{witting_time-domain_2016}. After attenuation by reflection from 2 glass wedges, the pulses are overlapped spatially and temporally with a bandpass-filtered portion of the residual pump pulse from the OPA in a 10~$\muup$m thick $\betaup$-barium borate (BBO) crystal. With a gate pulse at 810~nm, the sum-frequency signal lies between 240~nm and 650~nm, well within the detection range of a silicon-based detector. Using a separate beam rather than deriving the gate pulse from the HCF output has the additional advantage that weak pulses can be measured. An aperture attenuates the gate pulse and makes its focal spot larger than that of the HCF output, reducing the influence of pointing fluctuations.

The BBO crystal is cut for type-II (o-e-e) phase-matching ($\theta=46^\circ$, $\phi=30^\circ$), resulting in a phase-matching window that extends from 350~nm to beyond 3000~nm (see Supplementary Fig.~S5). To allow phase-matching, the polarization of the gate pulse is rotated by $90^\circ$ by a half-wave plate. After the interaction in the crystal, the signal beam is isolated by a combination of a Rochon prism and an aperture and then re-focused onto the entrance slit of a spectrometer (StellarNet Black Comet SR). Due to the non-collinear geometry, different spectral components of the upconverted signal travel in different directions; accurate re-imaging of the focus in the crossing plane is therefore critical. To achieve this, the spectrometer is placed in the tangential focus of an off-axis spherical mirror. A detailed sketch is shown in Supplementary Fig.~S5.

The TDP traces, obtained by scanning the delay of the gate pulse and recording spectra, are analysed using 200 iterations of the regularized ptychographic iterative engine \cite{maiden_further_2017}. To account for the effects of varying phase-matching efficiency, inherent frequency scaling of the nonlinearity, and the reflectivity of the components in the setup, we correct the trace using the frequency marginal as calculated from the measured spectra of the gate and unknown pulse. The shape of the pulses exiting the HCF is reconstructed by numerically back-propagating the retrieved field, taking into account the exit window of the HCF system as well as the path through argon and air.

\section*{Acknowledgements}
This work was funded by the European Research Council (ERC) under the European Union's Horizon 2020 research and innovation program: Starting Grant agreement HISOL, No. 679649.

\bibliography{bibliography}

\end{document}